\documentclass[aps,prl,superscriptaddress,twocolumn]{revtex4}
\usepackage{graphicx}
\usepackage{epstopdf}
\usepackage{bm}
\usepackage{amssymb}
\usepackage{color}

\begin{document}

\title{Magnetism in
 parent Fe-chalcogenides: quantum fluctuations select a plaquette order}

\author{Samuel Ducatman, Natalia  B. Perkins, Andrey Chubukov}
\affiliation{Department of Physics, University of Wisconsin-Madison, Madison, WI 53706, USA}
\date{today}%
\begin{abstract}
We analyze magnetic order in iron-chalcogenide Fe$_{1+y}$Te -- the parent compound of
 high-temperature superconductor Fe$_{1+y}$Te$_{1-x}$Se$_x$.
  Neutron scattering experiments show that magnetic order in this material
 contains components with momentum $Q_1=(\pi/2, \pi/2)$ and $Q_2 =(\pi/2, -\pi/2)$ in Fe-only Brillouin zone.
 The actual spin order depends on the
 interplay between these two
  components.
  Previous works argued that spin order is a single-$Q$ state (either $Q_1$ or $Q_2$). Such an order breaks rotational $C_4$ symmetry and
 order spins into a double diagonal  stripe.
  We show that quantum fluctuations actually select another order -- a double $Q$  plaquette  state
  with equal weight of $Q_1$ and $Q_2$ components,
   which preserves $C_4$ symmetry but breaks $Z_4$ translational symmetry. We argue that the plaquette state is consistent with recent neutron scattering experiments on Fe$_{1+y}$Te.
  \end{abstract}
\maketitle

{\it Introduction.} The analysis of magnetism in parent compounds of iron-based superconductors (FeSCs)
 is an integral part of the program to understand the origin of superconductivity in these materials~\cite{Ishida09,Johannes09,Paglione09,Qazilbush09,Chubukov09,Mazin09,Stanev08,Cvetkovich09,Eremin10,Johnston10,Peter11,
 Chubukov12}.
 Parent compounds of Fe-pnictides
 are moderately
  correlated metals, whose  resistivity increases with increasing $T$,  and the electronic structure is at least qualitatively consistent with that of free electrons on a lattice~\cite{basov,Chubukov09}.  Magnetic order in such systems can be reasonably well understood
  within itinerant scenario~\cite{Stanev08,Cvetkovich09,Vishwanath08,Eremin10}
   due to enhancement of free-electron susceptibility at momenta connecting hole and electron Fermi surfaces (FSs).
  The locations  of the FSs select two possible momenta for the order -- $(0,\pi)$ and $(\pi,0)$--
   in the Fe-only Brillouin zone (BZ).
   Electron-electron
    interaction and the shape of the FSs further reduce the ground state manifold to single-momentum states
    with either $(0,\pi)$ or $(\pi,0)$, but not their mixture
    ~\cite{Eremin10}).
     In each of these two states spins are ordered in a stripe fashion -- ferromagnetically along one direction in 2D Fe-plane
 and antiferromagnetically in the other. Such an order breaks $C_4$ lattice rotational symmetry and causes pre-emptive spin-nematic order~\cite{Fernandes12}. The same magnetic order is selected in the strong coupling approach, which assumes that the system is not far from Mott transition, and magnetic properties are reasonably well described by $J_1-J_2$ model with nearest and second-nearest neighbor spin exchange~\cite{localized,mila}.
 The actual coupling in Fe-pnicties is neither truly small nor strong enough to cause Mott insulating behavior~\cite{basov},
  which makes it extremely useful that the two descriptions agree.
    Upon doping, long-range order is lost, but magnetic fluctuations evolve smoothly and remain
    peaked at or near $(0,\pi)$ and $(\pi,0)$ even beyond optimal doping~\cite{ray}.
\begin{figure}
\includegraphics[width=0.45\columnwidth]{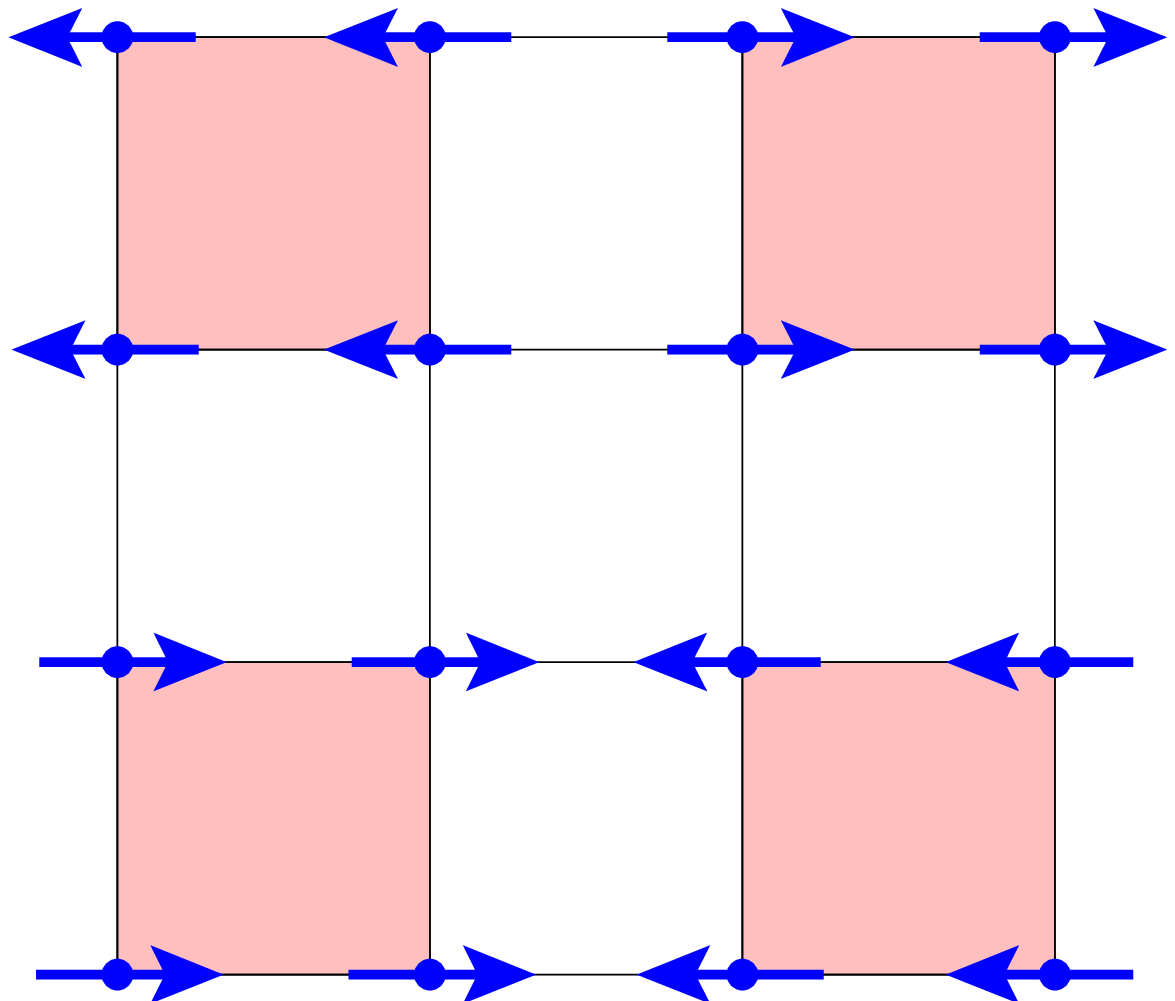}
\includegraphics[width=0.45\columnwidth]{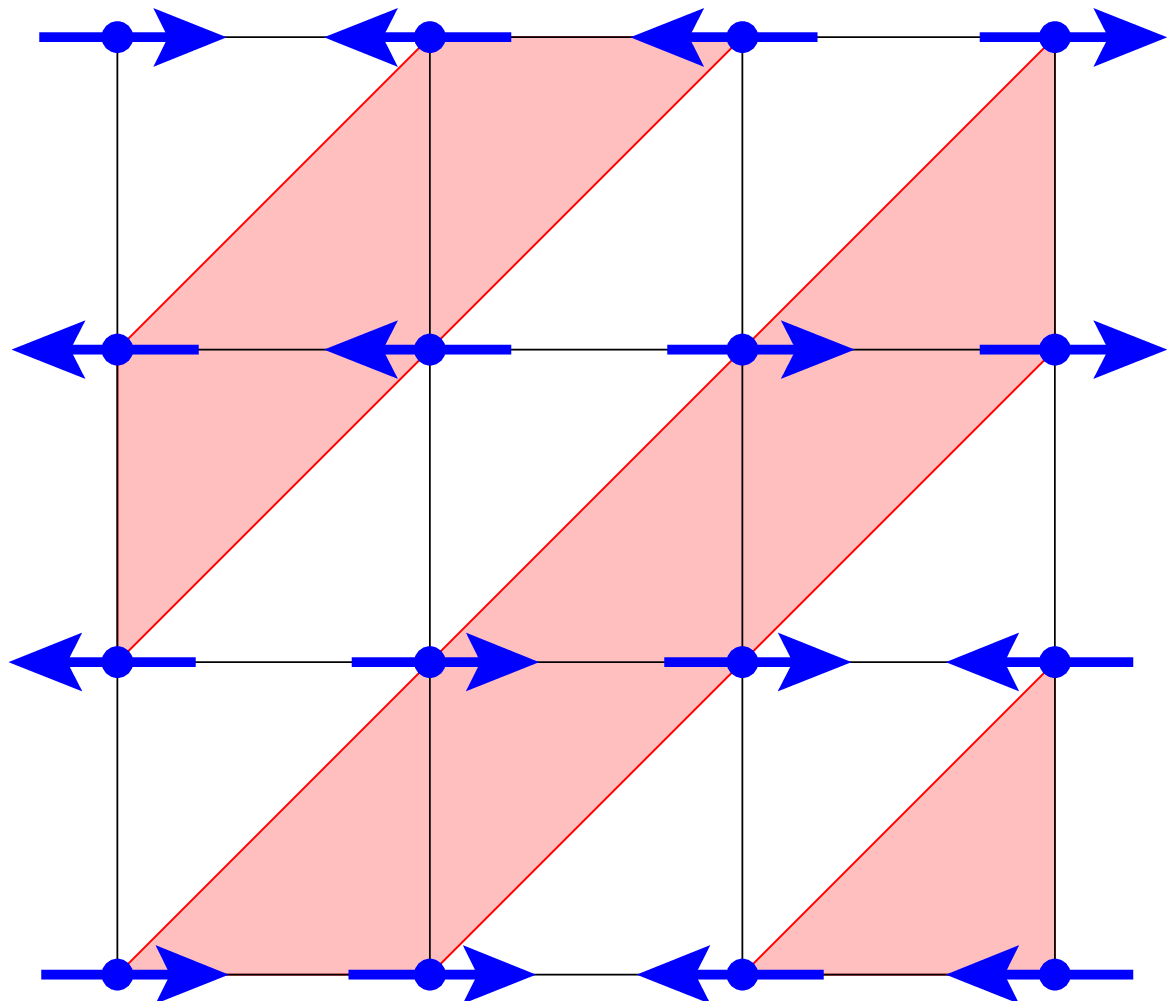}
\caption{
 The two possible collinear configurations for the
 $J_1-J_2-J_3$
  model: (a) orthogonal double stripe (ODS) and
(b) diagonal double stripe (DDS).
}
\label{fig:fig-stripes}
\end{figure}

There  is
 one family
 of FeSCs - 11 Fe-chalcogenides Fe$_{1+y}$Te$_{1-x}$Se$_x$,
  in which smooth evolution between parent and optimally doped compounds does not hold.
 Magnetism in these materials
  changes considerably between $x=0$ and $x \sim 0.5$, where the
 $T_c$ is the largest. Near optimal doping magnetic fluctuations are peaked at or near $(0,\pi)$ and $(\pi,0)$,
 as in Fe-pnictides,
 while
  magnetic order  in a parent  compound Fe$_{1+y}$Te  has very
   different momenta
   $\pm (\pi/2, \pm \pi/2)$~\cite{Li09,Liu10,Lipscombe11,Zaliznyak11,Zaliznyak12}.
     Upon doping, the spectral weight at $\pm (\pi/2, \pm \pi/2)$ decreases,
     and the spectral weight at
   $(0,\pi)$ and $(\pi,0)$ increases~\cite{Liu10}.
     The transport properties of Fe$_{1+y}$Te are also quite different from those
     of  parent compounds of Fe-pnictides:
the resistivity, $\rho (T)$, of Fe$_{1+y}$Te   does not show a prominent increase with increasing $T$, but instead remains flat and even  shows a small increase as  $T$ decreases~\cite{Mizuguchi10}.   ARPES studies of Fe$_{1+y}$Te  show that low-energy spectra are very broad~\cite{Xia09}, consistent with the notion that electrons are not propagating. These observations lead several groups to suggest that
 parent Fe-chalcogenides are more correlated than
 parent Fe-pnictides, and  magnetism  in Fe$_{1+y}$Te can be  understood
 by assuming that electrons are
"almost"
 localized and interact magnetically via a Heisenberg
   exchange~\cite{Subedi08,Ma09,Yu11,Hu12}.  This scenario is in line with a more generic idea~\cite{Si08,Zhao08,Yao2008} that in any FeSc,  a certain percentage of electronic states
     are localized and phase separated from itinerant electrons, and the percentage of localized states
     varies between different materials.
      An alternative scenario for FeTe,
       which we don't discuss here,
        is  orbital order~\cite{orbital}

In this communication we apply the localized electron scenario to Fe$_{1+y}$Te
 and verify whether the observed commensurate $\pm (\pi/2, \pm \pi/2)$ order can be obtained in a  Heisenberg model with exchange interactions up to third neighbors.
  Classically, $\pm (\pi/2, \pm \pi/2)$ order is unstable
   with respect to a spiral order for any non-zero first neighbor exchange, unless one artificially  breaks $C_4$ symmetry and sets interactions to be spatially anisotropic~\cite{Zhao08,Lipscombe11}. We analyze the
   isotropic quantum Heisenberg model and show that quantum fluctuations do stabilize a commensurate $\pm (\pi/2, \pm \pi/2)$ order in some range of parameters. However, this stabilization does not
  uniquely determine spin configuration
   as a generic $\pm (\pi/2, \pm \pi/2)$ order is a superposition of
   two
   different $Q-$vectors: ${\bf Q}_1 =(\pi/2, -\pi/2)$, and ${\bf Q}_2=(\pi/2, \pi/2)$:
    ${\bf S}({\bf r}) = {\bf \Delta}_{1} \cos {{\bf Q}_1 {\bf r}} +  {\bf \Delta}^{'}_{1} \sin {{\bf Q}_1 {\bf r}}
    + {\bf \Delta}_2 \cos {{\bf Q}_2 {\bf r}} + {\bf \Delta}^{'}_2 \sin {{\bf Q}_2 {\bf r}}$, with $|{\bf \Delta}_i| = |{\bf \Delta}^{'}_i| = \Delta$ and  ${\bf \Delta}_1 \cdot {\bf \Delta}_2 =
     {\bf \Delta}^{'}_1 \cdot {\bf \Delta}^{'}_2 =0$.
 In Fig. \ref{fig:fig-stripes}  we show two  prototypical commensurate spin configurations --
  a
  single $Q$
   bi-collinear spin order (${\bf \Delta}_1 = {\bf \Delta}^{'}_1 = {\bf \Delta}$,  ${\bf \Delta}_2 = {\bf \Delta}^{'}_2=0$), which
  breaks $C_4$, and a double $Q$
   plaquette order (${\bf \Delta}^{'}_1 = {\bf \Delta}_2 = {\bf \Delta}$,  ${\bf \Delta}_1 = {\bf \Delta}^{'}_2=0$),
   which preserves $C_4$ symmetry,
    but breaks $Z_4$ translational symmetry (four equivalent plaquette states are obtained by moving a black square
    in Fig. \ref{fig:fig-stripes}a by one lattice site in either $x$ or $y$ direction).
  Bi-collinear spin order is often called diagonal double stripe (DDS), and plaquette order is called orthogonal double stripe (ODS),
  and we use these notations below. The real-space configuration for both orders is "up-up-down-down"  along $x$ and $y$ directions.

 Most of previous theoretical and experimental works assumed
that the commensurate order is  DDS~\cite{Johnston10}
 and studied in detail the feedback from this order on electrons~\cite{Lipscombe11}.
We argue that
 quantum fluctuations of spins actually select ODS order as a stable collinear state
  for weak but finite nearest-neighbor exchange $J_1$,   while
   DDS  state is  unstable for any non-zero $J_1$.
    The DDS and the ODS orders have qualitatively different forms of the static structure factor $S(q)$ (two peaks vs four peaks),
   but this is difficult to detect in real materials because of domains.
   The authors of~\cite{Zaliznyak11} however argued that form of $S(q)$ in a paramagnetic phase allows one to distinguish between DDS and ODS, even in the presence of domains,
    and found that their results are consistent with strong ODS fluctuations.
  Another argument in favor of the $C_4$ preserving ODS spin order is the absence of orthogonal distortion in  Fe$_{1+y}$Te -- there is a monoclinic distortion below $T_N$, but this does not break rotational
  in-plane $C_4$ symmetry.
   There is also numerical evidence -- ODS  order has been found in exact diagonalization studies of $J_1-J_2-J_3$ model on clusters up to 36 spins \cite{Sindzingre09}.
   The same ODS order has been found in
   the mean-field studies of the $t-J$ model  in another Fe-chalcogenide K$_{0.8}$Fe$_{1.6}$Se$_2$~\cite{graf}.
  \begin{figure}
\includegraphics[width=0.6\columnwidth]{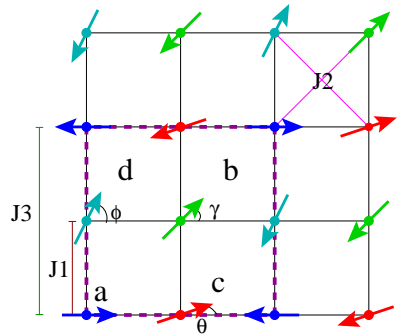}
\caption{
Spin order in the classical $J_1-J_2-J_3$ model at $J_1=0$.
Classically degenerate configurations form four sublattices, labeled  as $a,b,c$, and $d$.
 A configuration with arbitrary $\gamma$,  $\theta$, and  $\phi$ is a ground state.
 In our notations, sublattice spins are ${\bf \Delta}_1 + {\bf \Delta}_2$,  ${\bf \Delta}_1 - {\bf \Delta}_2$, ${\bf \Delta}^{'}_1 - {\bf \Delta}^{'}_2$, and ${\bf \Delta}^{'}_1 + {\bf \Delta}^{'}_2$, respectively.}
\label{fig:fig2}
\end{figure}

{\it Model.} We follow earlier works and model magnetic interactions in Fe$_{1+y}$Te by a $J_1-J_2-J_3$
 Heisenberg model
 ~\cite{Ma09,Sindzingre09,Reuter11,Yu11}:
\begin{equation}
H = \sum_{n=1}^3J_n \sum_{\langle ij\rangle} \vec{S}_i\cdot \vec{S}_{i+n}
\label{a_1}
\end{equation}
where $J_1, J_2$, and $J_3$ are antiferromagnetic exchange couplings between first-, second-, and third-nearest neighbors.  For Fe$_{1+y}$Te the values of $J_1, J_2$, and $J_3$ have been estimated in~\cite{Ma09}
 and found to be in the range
  $J_3>\frac{J_2}{2} \gg J_1$.
 In this limit, the  classical ground state of (\ref{a_1})
 is a spiral with the pitch vector $\mathbf Q = (\pm q, \pm q)$,  where $q=\arccos (\frac{-J_1}{2J_2+4J_3})$~\cite{Mambrini2006}.
 At $J_1 = 0$, the model has an extensive degeneracy,
  and any order with momentum $\pm (\pi/2, \pm \pi/2)$ is the classical ground state, including
   DDS, ODS, and an infinite number of other four-sublattice states
   (Fig. \ref{fig:fig2}).

   We consider here what happens in the quantum model,
  at a finite $J_1$.
    We show that the
   ODS state is unambiguously selected by quantum fluctuations
     to be the ground state in some range of $J_1$, before a  spiral order sets in.
      Our key reasoning is that only
       some  classically
    degenerate ground states at $J_1 =0$ are degenerate by symmetry;
     others are "accidentally degenerate".
    The situation is quite similar to the one in the well-known  $J_1-J_2$ model at $J_2 > J_1/2$~\cite{J1J2}.
    We argue that quantum fluctuations lift accidental degeneracies  and gap out some of the
    spin-wave modes which in the classical limit
    become unstable (imaginary) at $J_1 \neq 0$. For the
     DDS state the lifting of the accidental degeneracies does not help,
     as the modes which become
 unstable at a finite $J_1$ are the true Goldstone modes at $J_1=0$.
  On the other hand, for ODS state classically unstable modes are  accidental zero modes at $J_1=0$, and quantum fluctuations
  lift the energies of these modes to finite values,
   making ODS the state stable  in a finite range of $J_1$. We verified that ODS state
   is indeed the ground state in this range.

    {\it Large-S spin-wave calculations.} We consider large value of spin $S$ and study the role of quantum fluctuations within $1/S$ expansion.
    The computational steps are presented in~\cite{suppl}.
    For $J_1=0$, spins on even and odd sites form two non-interacting sublattices, each described by $J_2-J_3$ model.  This
     model is identical to $"J_1-J_2"$ model, with diagonal hopping $J_2$ playing the role of $"J_1"$ and third-neighbor hopping
     $J_3$ playing the role of $"J_2"$. One can use this analogy and  borrow the results of the quantum analysis of $"J_1-J_2"$ model~\cite{J1J2}.
     For $J_3 > J_2/2$ (which holds in Fe$_{1+y}$Te), quantum fluctuations select stripe configurations within each sublattice, i.e. the angle $\gamma$ in Fig.\ref{fig:fig2} is locked at $\gamma =0$ or $\gamma = \pi$,  and the angle $\theta$ is locked at $\theta =\phi$ or $\theta = \pi + \phi$.
     The states with $\gamma=0$ and $\gamma = \pi$ are equivalent up to an interchange of $X$ and $Y$ directions, and below we set $\gamma =0$.
 The
      collinear DDS and ODS states belong
       to the manifold of selected states and correspond to different locking of the angle $\phi$ between the nearest-neighbor spins:
         DDS state corresponds to $\phi =0, \theta =\pi$ or $\phi = \pi, \theta =0$, while ODS corresponds to $\phi =\theta =0$ or $\phi = \theta =\pi$.

 To analyze  whether a generic state selected by quantum fluctuations at $J_1 =0$
 remains stable at a finite value of $J_1$,  we  need to know
  its excitation spectrum.
 At  $J_1 =0$,
  spins on even and odd sites are decoupled,  each sublattice is
  described by its own
  bose field ($\alpha_\mathbf{k}$ for even sites and $\beta_\mathbf{k}$ for odd sites), and spin-wave
 excitations are described by
 \begin{equation}
 H_{sw} = S (\Omega_{\alpha \mathbf{k}} \alpha^\dagger_\mathbf{k} \alpha_\mathbf{k} + \Omega_{\beta \mathbf{k}} \beta^\dagger_\mathbf{k}\beta_\mathbf{k}),
 \label{eq:sw}
 \end{equation}
  The classical spin-wave spectrum is the same for all selected states
\begin{eqnarray}
&&\Omega_{\mathbf{k}} = S(A^2_\mathbf{k} - B^2_\mathbf{k})^{1/2},~A_\mathbf{k} = 4J_3 + 2J_2 \cos(k_x+k_y),  \nonumber \\
&&B_\mathbf{k} = 2J_2 (\cos 2k_x+\cos 2k_y) + 2J_2 \cos(k_x-k_y).
\label{a_2}
\end{eqnarray}
This  spectrum contains  nodes at $\pm (\pi/2, \pm \pi/2)$, but some of them are not symmetry-related and
 are lifted by quantum fluctuations.
For the sublattice made of even sites, the order has momentum $\pm (\pi/2, -\pi/2)$ (Fig. \ref{fig:fig2} b), hence the true nodes are
 located only at these momenta, while the ones at $\pm (\pi/2,\pi/2)$ must be lifted.
 For the sublattice made out of spins at odd sites, the order has momentum $\pm (\pi/2,\pi/2)$ if we take $\theta = \phi$, like in the ODS, and
 $\pm (\pi/2, -\pi/2)$ if we take $\theta = \pi + \phi$, like in the DDS. Quantum fluctuations then must lift the nodes at
 $\pm (\pi/2,-\pi/2)$  and at $\pm (\pi/2,\pi/2)$ for the ODS and the DDS  state,
  respectively.
   We computed quantum corrections to the spectrum in Eq. (\ref{a_2})
  within  perturbation theory to order $1/S$
    and indeed found that
 accidental nodes are lifted by quantum fluctuations and only true Goldstone
modes remain~\cite{suppl}.

 We next set $J_1$ to be small but finite and consider which of stripe states, if any, remain stable.
The qualitative reasoning is the following: a non-zero $J_1$ couples the two sublattices and adds to the Hamiltonian (\ref{eq:sw}) the terms in the form
  $\alpha^\dagger_\mathbf{k} \beta_\mathbf{k}$ and $\alpha_\mathbf{k} \beta_\mathbf{k}$.  For the DDS state
  (or, more accurately, for the DDS family of states as we keep $\phi$ as a parameter)
   the stripes on even and odd sites are directed parallel to each other, and the dispersions of $\alpha_\mathbf{k}$ and $\beta_\mathbf{k}$ fields are identical, including $O(1/S)$ terms. The two dispersions are then gapless at the same momenta
   $\mathbf{k} = \pm (\pi/2, -\pi/2)$.  Around these $\mathbf{k}$ points,   the perturbation theory in $J_1$ is singular,
     as there is no symmetry requirement which would force the coupling to vanish at $\pm (\pi/2, -\pi/2)$. As a result, the excitations become purely imaginary close enough to $\pm (\pi/2, -\pi/2)$, which implies that
   the DDS states are unstable at any non-zero $J_1$.
On the other hand, for the ODS family of states, the dispersions $\Omega^{\alpha}_{ \mathbf{k}}$ and $\Omega^{\beta}_{\mathbf{k}}$ have nodes at different momenta, $\pm (\pi/2, -\pi/2)$ and $\pm (\pi/2, \pi/2)$, respectively. Because of this disparity,
      perturbation  theory near either $\pm (\pi/2, -\pi/2)$ or $\pm (\pi/2, \pi/2)$ is not singular, and corrections in $J_1$ only gradually shift the values of spin-wave velocities
 thus keeping ODS states stable.

  We verified this reasoning by
 explicit calculations.  We first obtained the $J_1$-induced interaction in terms of
  the original Holstein-Primakoff bosons and then re-expressed it in terms of
  $\alpha_\mathbf{k}$ and $\beta_\mathbf{k}$ bosons from Eq. (\ref{eq:sw}),
    which are related to the original ones by Bogoliubov transformation.  The $u_\mathbf{k}v_\mathbf{k}$-coefficients of this transformation dress up  the interaction terms.
  For the DDS states, expanding the Hamltonian near  the true Goldstone points at $(\pi/2, -\pi/2)$ as $\mathbf{k} = (\pi/2, -\pi/2) + \mathbf{\tilde k}$
  we obtain  $H_{DDS} = H_{sw} + \delta H_{DDS}$, where
$H_{sw}$ is given by (\ref{eq:sw}) with
\begin{eqnarray}
\Omega^\alpha_{\mathbf{\tilde k}} = \Omega^\beta_{ \mathbf{\tilde k}}
 \approx 4S \sqrt{J_3 (2J_3 + J_2)}({\tilde k}^2_x+{\tilde k}^2_y -2a {\tilde k}_x {\tilde k}_y)^{1/2},
 \label{a_3_1}
 \end{eqnarray}
  where $a = \frac{J_2}{2J_3} <1$,  and
 \begin{eqnarray}
\delta H_{DDS} = \Delta^{DDS}{\mathbf{\tilde k}} (\alpha_{\mathbf{\tilde k}}^\dagger \beta_{\mathbf{\tilde k}} + \alpha_\mathbf{\tilde k} \beta_\mathbf{-{\tilde k}}  + h.c)
\label{a_3}
\end{eqnarray}
where
\begin{eqnarray}
\Delta^{DDS}_{\mathbf{\tilde k}}= \frac{J_1 S}{2} \left(\frac{2J_3+J_2}{J_3}\right)^{1/2} \frac{{\tilde k}_y - {\tilde k}_x}{({\tilde k}^2_x+{\tilde k}^2_y -2a {\tilde k}_x {\tilde k}_y)^{1/2}}
\label{a_4}
\end{eqnarray}
The coupling term remains finite when ${\tilde k}_{x,y}$ tends to zero, except for special directions.
Diagonalizing (\ref{a_3}) we find that at low enough ${\tilde k}$ one of the two solutions is
$E^2_{\tilde k} \approx - 2\Omega^{\alpha(\beta)}_{\mathbf{\tilde k}} \Delta^{DDS}_{\mathbf{\tilde k}}$. A negative $E^2_{\tilde k}$ implies that fluctuations around a DDS  state
 grow exponentially with time and make this
 family of states unstable.

 For the ODS states the situation is different because near any of the points $\pm (\pi/2, -\pi/2)$ or $\pm (\pi/2,\pi/2)$, the zero in
  one of the spin-wave branches is  lifted by quantum fluctuations.  For example, near $(-\pi/2,\pi/2)$  expanding  of the Hamiltonian again gives
  $H_{ODS} = H_{sw} + \delta H_{ODS}$,  however now only $\Omega^{\alpha}_{\mathbf{\tilde k}}$ is gapless,
     while
  $\Omega^{\beta}_{\mathbf{\tilde k}}$ is gapped with the gap of the order $1/S$.
  The interaction term $\delta H_{ODS}$ has the same form as in (\ref{a_3}), but with
  \begin{equation}
  \Delta^{ODS}_{{\mathbf{\tilde k}}} = 2 J_1 S^2 (2J_3+J_2) \frac{{\tilde k}_y - {\tilde k}_x}{(\Omega^\alpha_{\tilde k}
  \Omega^\beta_{\tilde k})^{1/2}}
   = O\left(J_1 S^{3/2} |{\tilde k}|^{1/2}\right).
  \label{a_5}
  \end{equation}
  Diagonalizing $H_{ODS}$ we find two solutions,
  \begin{eqnarray}
  E^2_{1,2} &=& \frac{1}{2}\Bigl(
  (\Omega^{\alpha}_{\tilde k})^2 + (\Omega^\beta_{\tilde k})^2 \\\nonumber&&\pm
  \sqrt{((\Omega^\alpha_{\tilde k})^2 - (\Omega^\beta_{\tilde k})^2)^2 + 16 (\Delta^{ODS}_{\mathbf{\tilde k}})^2 \Omega^\alpha_{\tilde k}. \Omega^\beta_{\tilde k}}\Bigr).
  \end{eqnarray}
   One of the solutions
   is gapped to order $1/S$, the other
    is linear in ${\tilde k}$
  with a stiffness which differs from its value at $J_1=0$  by
  $O(J_1 S/J_3)$.
   We see that the ODS states are stable (for any $\phi$)  as long as $J_1S/J_3$ is small.

  On a more careful look, we find that the ODS spin order allows for $J_1-$induced umklapp processes, which also
    renormalize the dispersions of the ODS states. Indeed, because
    ODS state breaks $Z_4$ translational symmetry,
    the $J_1$ interaction contains not only the terms at zero transferred momentum, as in  (\ref{a_3}), but also terms with momentum
     transfer in multiples of $\pi$ along each axis. Near $k = (\pi/2,-\pi/2)$, the most relevant of such umklapp terms is the one
     with momentum ${\bar Q} =(0,\pi)$, which connects
       a gapless $\alpha_{\tilde k}$ boson
        at  $(\pi/2, -\pi/2)$, and a gapless $\beta_{\tilde k}$ boson
        at $(\pi/2,\pi/2)$.
  However, because
   breaking of $Z_4$ is equivalent to breaking local inversion symmetry (a reflection around one column or one row in Fig. \ref{fig:fig-stripes}a),
     the umklapp vertices $\Delta^{U,ODS}_{\mathbf{\tilde k}}$ contain extra momentum gradient compared to non-umklapp vertices.  In explicit form, we find
at small ${\mathbf{\tilde k}} = \mathbf{k} - (\pi/2,-\pi/2)$,
  \begin{eqnarray}
  \Delta^{U,ODS}_{\mathbf{\tilde k}} = -i \frac{J_1}{4} \left(\frac{\Omega^{\alpha}_{\tilde k}
  \Omega^{\beta}_{{\tilde k} + {\bar Q}}}{4J^2_3-J^2_2}\right)^{1/2}\cos \phi~,
    \label{a_7}
  \end{eqnarray}
where  $\Omega^{\alpha}_{\tilde k}, \Omega^{\beta}_{\tilde k + {\bf Q}} = 4 S (J_3 (2J_3 \pm J_2))^{1/2}
   ({\tilde k}^2_x+{\tilde k}^2_y \mp 2   {\tilde k}_x {\tilde k}_y)^{1/2}$ and the angle $\phi$ specifies the spin order within  the ODS family  of states.
   We see that $\Delta^{U,ODS}_{\mathbf{\tilde k}}$
    scales linearly with ${\mathbf{\tilde k}}$, i.e.,  is of the same order as $\Omega^{\alpha \beta}_{\tilde k}$.
    We computed the corrections to
    spin-wave velocity and
    found that they
    scale as $J_1/\sqrt{4J^2_3-J^2_1}$, i.e., are small.
At the same time, we
 see from the Eq.(\ref{a_7}) that
  $\Delta^{U,ODS}_{\mathbf{\tilde k}}$ depends on the angle $\phi$. Respectively, the corrections to the ground state energy
 also depend on $\phi$ and
 should select which state within the ODS family
  has the lowest energy. The computation
  is straightforward and yields
 $\Delta E_{gr} = - A \cos^2 \phi$, with $A>0$.
 We see that the
 collinear ODS state, for which
 $\phi =0$ or $\pi$, is indeed the state with the lowest energy.

The outcome of our analysis is that
  the collinear ODS state remains stable and has the lowest energy within a family of similar states.  At small $J_1$, the ODS state has a finite stiffness towards fluctuations which tend to break collinear order in favor of a spiral one. The ODS state remains stable up to $J_1 \sim J_3/S$,  at larger $J_1$ the stiffness changes sign, and the system develops a spiral order.

{\it Experimental signatures of ODS state.}
Because the ODS state does not break $C_4$ translational symmetry, it does not cause a pre-emptive spin-nematic order, in contrast to
 parent compounds of other FeSCs~\cite{Fernandes12}. The data for Fe$_{1+y}$Te show that the system develops a monoclinic distortion below a certain $T$, but in-plane $C_4$ symmetry remains unbroken (it only breaks in  doped compounds Fe$_{1+y}$Te$_{1-x}$Se$_x$ with $x >0.5$ ~\cite{Mizuguchi10}).  The  unbroken $C_4$ symmetry in the ordered state  also manifests itself in the
 $C_4$ symmetry of the
 static structure factor $S(q)$ obtained  in neutron scattering experiments ~\cite{Zaliznyak11}. We computed $S(q)$  for both the DDS and the ODS states,  and we indeed found that
  the structure factor  for the ODS order has four identical peaks at $(\pm \pi/2, \pm \pi/2)$, while the structure factor for the DDS state has only two peaks at $(\pi/2, -\pi/2)$ and $(-\pi/2, \pi/2)$.
   While the observed four peaks are consistent with ODS, we caution that
   the  absence of  the anisotropy  in  the structure factor obtained in neutron scattering could be due to the twinning of the crystal. However, as the magnetic domain's structure of the crystal can
    be controlled using polarized neutrons,  the careful analysis of the neutron scattering  data might dissect   the contribution from different domains.
    The authors of Ref.~\cite{Zaliznyak11} made another argument that, even in a twinned crystal, the form of $S(q)$ throughout the Brillouin zone differentiates between strong DDS and ODS fluctuations, and argued that their data are more consistent with tendency towards ODS order.  This again agrees with our results.

{\it Summary.} In this communication we analyzed the type of magnetic order in Fe$_{1+y}$Te -- the parent compound in a family
 of Fe-chalcogenide superconductors.  The magnetic order in this material is different from  in other parent compounds of FeSCs  -- spins are ordered in up-up-down-down fashion (Fig. \ref{fig:fig2}).
 Experiments show~\cite{Mizuguchi10,Tamai10} that the tendency towards Mott physics is  stronger in Fe$_{1+y}$Te than in other parent compounds of FeSCs, suggesting that the magnetic order in Fe$_{1+y}$Te can be reasonably well understood within the localized scenario
  by solving the Heisenberg model with exchange interaction extending up to third neighbors~\cite{Ma09}.
    Several groups argued~\cite{Ma09,Li09,Lipscombe11,Fang12} that the ordered up-up-down-down spin configuration is diagonal double stripe. Such an order breaks $C_4$  lattice rotational symmetry. We argued, based on our analysis of quantum fluctuations in the Heisenberg model with first, second, and third-neighbor exchange, that such a state is unstable, but another
    up-up-down-down state -- the orthogonal double stripe, is stable and is the  ground state in some parameter range. This state
    (which is also called a plaquette state) breaks $Z_4$ translational symmetry, preserves $C_4$ symmetry, and does not cause
  orthorhombic distortion. Also, its structure factor has four equivalent peaks at $(\pm \pi/2, \pm \pi/2)$, in agreement with
  recent neutron scattering studies of Fe$_{1+y}$Te~\cite{Zaliznyak11}. An interesting issue that deserves further study is whether $Z_4$ translational symmetry can be broken before a true ODS spin order sets in, as it happens in other systems~\cite{chern}.

 \textit{Acknowledgement.} We acknowledge useful conversations with C. Batista, R. Fernandes, G-W. Chern,
 M. Graf, B. Lake, A. Nevidomskyy, J. Schmalian, N. Shannon, and I Zaliznyak.
 N.B.P. is supported by  NSF-DMR-0844115, A.V.C. is supported by
NSF-DMR-0906953.

\newpage
\begin{widetext}
\begin{center}

{ \large Supplemental Material }\\\vspace{0.3cm}

Samuel Ducatman, Natalia  B. Perkins, Andrey Chubukov\\\vspace{0.3cm}

Department of Physics, University of Wisconsin-Madison, Madison, WI 53706, USA
\end{center}
\end{widetext}

 In this Supplementary Material we provide details  of  our  analysis of how quantum fluctuations lift  accidental nodes
  in the spin-wave spectra
 of the sub-set of states selected by quantum fluctuations
   in the  $J_1-J_2-J_3$ model at $J_3 > J_2/2$ and $J_1=0$ (i.e., in the $J_2-J_3$ model with $J_3 > J_2/2$).
    The Hamiltonian for the model is given by Eq.~(1) in the main text.  At $J_1 =0$ spins on
   even and  odd sites do not interact with each other. It is therefore sufficient to consider
   only the sub-set made out of spins on, say, even sites.

 Classically degenerate ground states are shown in Fig. 2 of the main text.  For spins on even sites, any two-sublattice
 configuration with arbitrary angle $\gamma$ between spins in $a$ and $b$
  sublattices is the ground state. Like we said in the text, quantum fluctuations break this degeneracy and select the states at $\gamma =0$ or $\gamma = \pi$ as the only two ground states.  The state with $\gamma =0$ can be viewed as a spiral state with $Q_1 = (\pi/2, -\pi/2)$, and the $\gamma =\pi$ state corresponds to a spiral with $Q_2 = (\pi/2, \pi/2)$.
   Because only even cites are involved, configurations with $Q_i$ and $-Q_i$ are identical.

  The classical spectrum of either $Q_1$ or $Q_2$ state contains zero modes at corresponding $Q$, which must be there by Goldstone theorem, but also contains zero modes at "wrong Q"  (at $Q_2$ for the $Q_1$ spiral and vise versa). These last modes are not associated with symmetry breaking (i.e., are "accidental") and must be lifted by quantum fluctuations.  In the main text we stated that quantum fluctuations do gap accidental zero modes  and studied the consequences.  Here we show how this actually happens. For definiteness, we consider $Q_1$ configuration ($\gamma =0$).

The lifting of accidental zero modes can be studied within the original two-sublattice picture, with two different bosons describing fluctuations of spins in $a$ and $b$ sublattices. It is more convenient, however,
 to  perform a uniform rotation
  to a local reference frame in which the magnetic order is ferromagnetic and describe the excitation spectrum using just one
  Holstein-Primakoff  bose operator $a_i$:
 \begin{eqnarray}
    \label{eq:HP}\nonumber
    \begin{array}{l}
    S^z_i =S - a^{\dagger}_i a_i \\
    S^+_i =(2S - a^{\dagger}_i a_i)^{\frac{1}{2}}a_i~~~~~~~~~~(S.1) \\
    S^-_i = a^{\dagger}_i (2S -a^{\dagger}_ia_i)^{\frac{1}{2}}~,
    \end{array}
\end{eqnarray}

As it is customary to spin-wave analysis, we assume that $S$ is large.
Our goal is to obtain the excitation spectrum for $Q_1$ configuration including terms
 of order $1/S$, which describe quantum fluctuations.
Accordingly, we   substitute Eq.~(S.1) into
Hamiltonian (Eq.~(1)) and then expand  it to the quartic order in $a$ field.
 We obtain
 \begin{eqnarray}
    \label{eq:Hb}\nonumber
    H = E_0+S(H_2+H_4)~,~~~~~~~(S.2)
\end{eqnarray}
where $H_2$ and $H_4$ are the quadratic and the quartic terms, respectively.
In momentum space, we have
 \begin{eqnarray}
    \label{eq:H2}
\nonumber
H_2 =& \sum_\mathbf{k} \Bigl( A_\mathbf{k}a^{\dagger}_\mathbf{k}a_\mathbf{k}
 +\frac{B_\mathbf{k}}{2}(a_\mathbf{k}a_\mathbf{-k}+h.c.)
\Bigr ),~~(S.3)
\end{eqnarray}
where
\begin{eqnarray}\nonumber
A_\mathbf{k}&=& 4J_3 +2J_2\cos(k_x+k_y) ~~~~~~~~~~~~~~~(S.4)  \\\nonumber
B_\mathbf{k}&=& -2J_3(\cos(2k_x)+\cos(2k_y))-2J_2 \cos(k_x-k_y)
\end{eqnarray}
and ${\bf k}$ is defined in the first magnetic BZ.
The quartic  part of the large-S expansion   is  given by
\begin{eqnarray}\label{eq:H4}
&&H_4 = \frac{J_2}{8NS}
\sum_{\{\mathbf k_i\}}\delta\Bigl(\sum_i\mathbf k_i \Bigr)
\Bigl[  a^{\dagger}_{-{\mathbf k}_1}a_{{\mathbf k}_2}a_{{\mathbf k}_3}a_{{\mathbf k}_4}\cos(k_{4x}-k_{4y})
\nonumber\\\nonumber
&&
-a^{\dagger}_{-{\mathbf k}_1}a^{\dagger}_{-{\mathbf k}_2}a_{{\mathbf k}_3}a_{{\mathbf k}_4}
\cos(k_{4x}+k_{4y})
~~~~~~~~~~~~~~~~(S.5)
\\\nonumber
&&-2a^{\dagger}_{-{\mathbf k}_1}a^{\dagger}_{-{\mathbf k}_2}a_{{\mathbf k}_3}a_{{\mathbf k}_4}
\sin(k_{2x}+k_{4x})\sin(k_{2y}+k_{4y}) +h.c.\Bigr]
\\\nonumber
&& +\frac{J_3}{8NS}\sum_{\{\mathbf k_i\}}\delta\Bigl(\sum_i\mathbf k_i \Bigr)
\Bigl[
a^{\dagger}_{-{\mathbf k}_1} a_{{\mathbf k}_2} a_{{\mathbf k}_3}a_{{\mathbf k}_4}(\cos 2k_{4x}+\cos 2k_{4y})\\\nonumber
&&-a^{\dagger}_{-{\mathbf k}_1}a^{\dagger}_{-{\mathbf k}_2} a_{{\mathbf k}_3}a_{{\mathbf k}_4}(\cos 2(k_{2y}+k_{4y})+
        \cos 2(k_{2x}+k_{4x})) +h.c.\Bigr]
\end{eqnarray}

The quadratic Hamiltonian $H_2$ is diagonalized
 by introducing   the Bogoliubov
transformation, $\alpha_\mathbf{k} = u_\mathbf{k}a_\mathbf{k}-v_\mathbf{k}a^{\dagger}_\mathbf{-k}$,
where  the coherence factors $u_\mathbf{k}$ and $v_\mathbf{k}$ are determined by
 \begin{eqnarray}
\nonumber
u_\mathbf{k} &=& \frac{1}{2}\sqrt{\frac{A_\mathbf{k}+\Omega_{\mathbf{k}}}{\Omega_{\mathbf{k}}}}~, \\
v_\mathbf{k}&=& -\frac{\text{sign}B_\mathbf{k}}{2}\sqrt{\frac{A_\mathbf{k}-\Omega_{\mathbf{k}}}
{\Omega_{\mathbf{k}}}}\nonumber
\end{eqnarray}
 and
\begin{eqnarray}\nonumber
\Omega_{\mathbf{k}}= S\sqrt{A_\mathbf{k}^2 - B_\mathbf{k}^2}~.
\end{eqnarray}
The diagonalized Hamiltonian $H_2$ is  given by
\begin{eqnarray}\nonumber
H_2=E_{2} + \sum_\mathbf{k} \Omega_{\mathbf{k}}\alpha^\dagger_\mathbf{k} \alpha_\mathbf{k}~,~~~~~~~~(S.6)
\end{eqnarray}
where
\begin{eqnarray}
    \label{e2}
E_{2}=\frac{S}{2}\sum_{\mathbf k} (-A_\mathbf{k} + \Omega_{\mathbf{k}})~~~~~~~~~~~~(S.7)\nonumber
\end{eqnarray}
 is the  contribution to the ground state energy from non-interacting $a$-bosons

 The spin-wave dispersion  of non-interacting bosons is  $\Omega_{\mathbf{k}}$.
  At  $\mathbf{k} = \pm Q_1$ and $\mathbf{k} = \pm Q_2$, $A_{Q_{1(2)}}=B_{Q_{1(2)}}=4J_3 \pm 2J_2$, where  + and - signs are  for  $Q_1$ and $Q_2$, respectively.  As a result, $\Omega_{Q_1}=\Omega_{Q_2}=0$.  However, as we said, only the zero mode at $\mathbf{k} = \pm Q_1$ is the
   true Goldstone mode,  the other one is accidental and must be gapped by quantum fluctuations.
   To show this, we have to
  compute 1/S corrections  to the spectrum at these points.

\begin{figure}
\includegraphics[width=0.99\columnwidth]{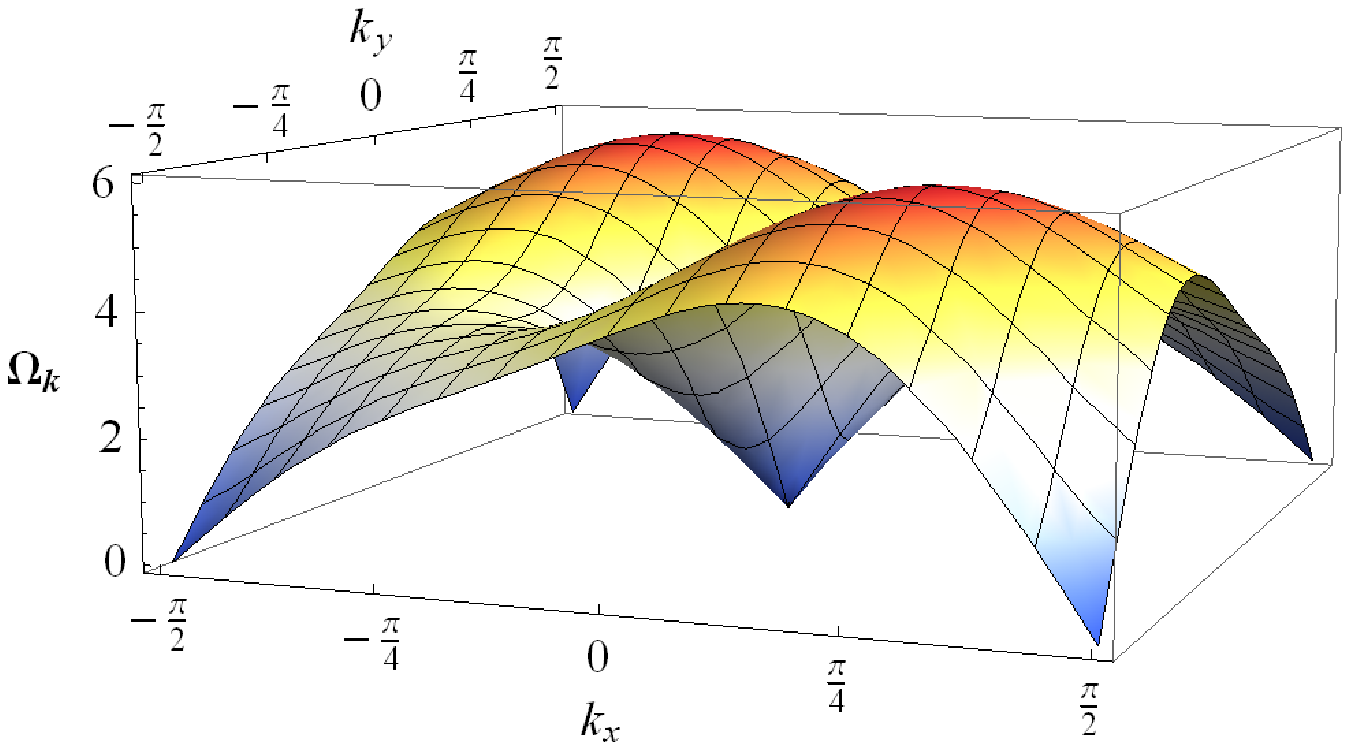}
\includegraphics[width=0.99\columnwidth]{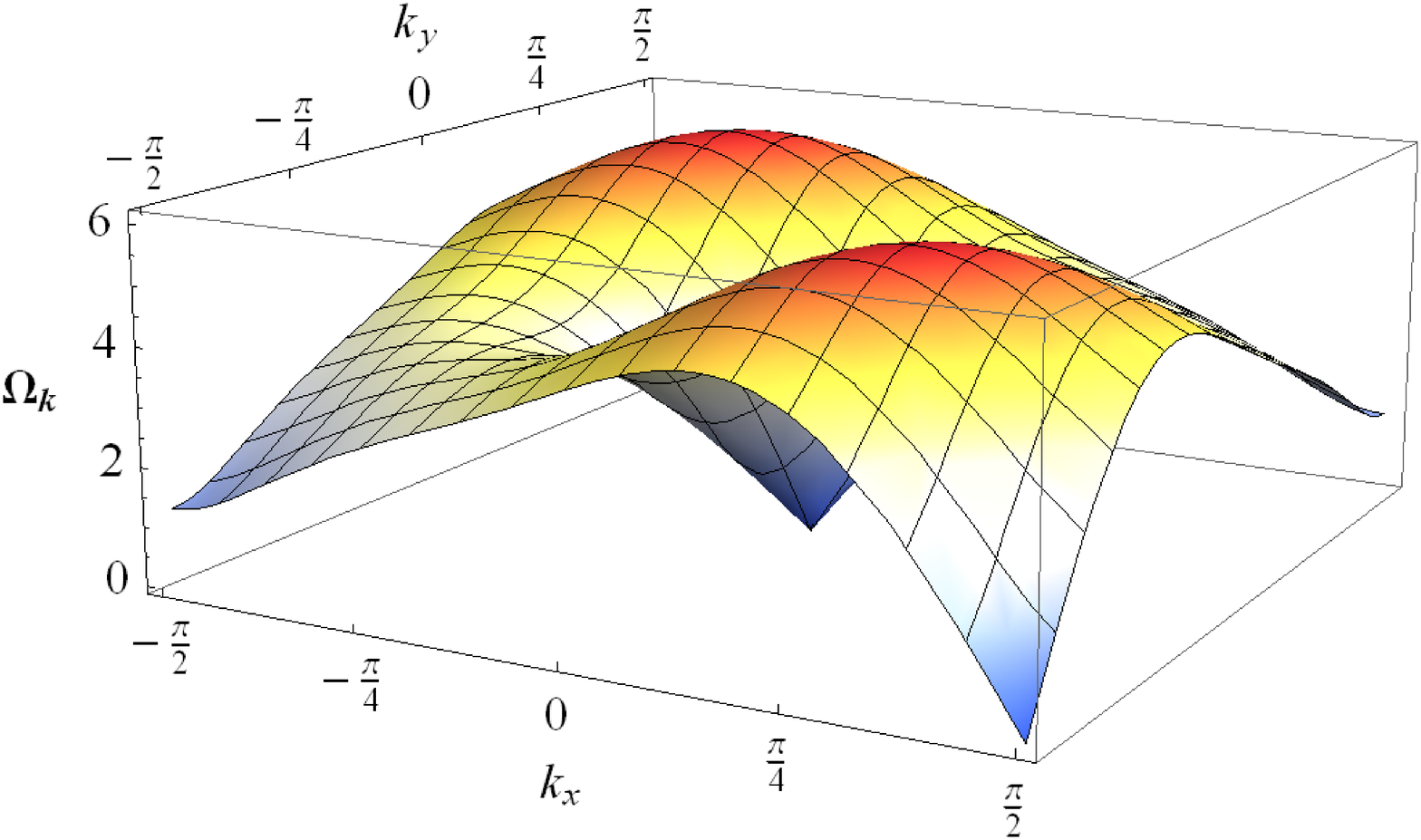}
\caption{Excitation spectrum for $Q_1$ spiral.
 Top panel: The spectrum of non-interacting bosons, Eq.(S.6)
 This spectrum contains zero modes at $\pm (\pi/2, \pi/2)$ and $\pm (\pi/2, -\pi/2)$
  Bottom panel: the spectrum
renormalized by  quantum fluctuations to order $1/S$.
 The true Goldstone modes at
$\pm\mathbf{Q_{1}}=\pm(\frac{\pi}{2},-\frac{\pi}{2})$ remain, but  the  accidental
zeroes at $\pm \mathbf{Q}_2=\pm(\frac{\pi}{2},\frac{\pi}{2})$   are
removed  by quantum fluctuations.}
\label{fig:spectrum}
\end{figure}

The $1/S$ contribution to $\Omega_{\mathbf{k}}$ is obtained by evaluating the first-order correction from $H_4$ what amounts to decoupling
   of $H_4$ into the product of two pairs of $a-$ bosons and replacing one pair by its average value for non-interacting bosons.
   We first define the averages
\begin{eqnarray}
K_{{\mathbf q}} &=& \langle a_{\mathbf q}^{\dagger} a_{\mathbf q} \rangle  = v_{\mathbf q}^2 ~,~~~~~~(S.8)\nonumber\\
L_{{\mathbf q}} &=& \langle a_{\mathbf q} a_{-{\mathbf q}} \rangle = \langle a_{-\mathbf q}^{\dagger} a_{{\mathbf q}}^{\dagger}  \rangle = u_{\mathbf q}v_{\mathbf q} ~.\nonumber
\end{eqnarray}
The decoupled $H_4$ is
\begin{eqnarray}
{\bar H}_4 &=&\frac{1}{8NS} \sum_{\mathbf k,q}
\Bigl(a^{\dagger}_{\mathbf k} a_{\mathbf k} F_1({\mathbf k},{\mathbf q})+\nonumber\\&&
(a_{\mathbf k} a_{-{\mathbf k}} +a_{-\mathbf k}^{\dagger} a_{{\mathbf k}}^{\dagger}  )\frac{F_2({\mathbf k},{\mathbf q})}{2}\Bigr),~(S.9)\nonumber
\end{eqnarray}
where
\begin{eqnarray}
&&F_1({\mathbf k},{\mathbf q}) = -2J_3\Bigl( 4K_{{\mathbf q}} (\cos^2(k_x-q_x)+\cos^2(k_y-q_y)) \nonumber\\
 &&- L_{{\mathbf q}} (2(\cos 2q_x+\cos 2q_y)+\cos 2k_x+\cos 2k_y)\Bigr) \nonumber\\
&&- 2J_2 \Bigl(K_{{\mathbf q}}(4\sin (k_x-q_x)\sin (k_y-q_y)\nonumber\\&&
+2\cos (k_x+k_y)+2\cos (q_x+q_y))\nonumber\\
&&-L_{{\mathbf q}} (2\cos(q_x-q_y)+\cos(k_x-k_y))\Bigr),\nonumber\\
&&F_2({\mathbf k},{\mathbf q}) = 2J_3
\Bigl( K_{\mathbf q}(
\cos 2q_x+\cos 2q_y \nonumber\\
&&+2(\cos 2k_x +\cos 2k_y))\nonumber\\
&&-2L_{\mathbf q} (\cos 2(k_x-q_x)+\cos2(k_y-q_y))\Bigr )\nonumber\\
&&+2J_2 \Bigl( K_{{\mathbf q}} (\cos(q_x-q_y)+2\cos(k_x-k_y))\nonumber\\
&&+L_{\mathbf q}(-4\sin (k_x-q_x)\sin (k_y-q_y)\nonumber\\&&
-\cos (k_x+k_y)-\cos (q_x+q_y))\Bigr)
\nonumber
\end{eqnarray}

Combining the  decoupled $H_4$ with $H_2$ expressed in terms of original $a$-operators we obtain
\begin{eqnarray}\label{h4deca}
    &&H_2 + {\bar H}_4
    =   \sum_{\mathbf k} \Bigl[ (A_\mathbf{k}+\frac{A^{(4)}_\mathbf{k}}{S})a^{\dagger}_\mathbf{k} a_\mathbf{k}
     + \nonumber\\&&
  \frac{1}{2}(B_\mathbf{k}+\frac{B^{(4)}_\mathbf{k}}{S})(a_\mathbf{k}a_\mathbf{-k}+h.c.)\Bigr],~~~~(S.10)\nonumber
\end{eqnarray}
where
\begin{eqnarray}
A^{(4)}_\mathbf{k}&=& \frac{1}{8N}\sum_\mathbf{q} F_1({\mathbf k},{\mathbf q}),\nonumber\\
B^{(4)}_\mathbf{k} &=& \frac{1}{8N}\sum_\mathbf{q} F_2({\mathbf k},{\mathbf q})\nonumber
\end{eqnarray}

At $\mathbf{k} = Q_1$ and $\mathbf{k}= Q_2$ we have
\begin{eqnarray}
A^{(4)}_\mathbf{Q_{1(2)}}&=& \frac{-1}{4N}\sum_\mathbf{q} J_3\Bigl( 4K_{{\mathbf q}} (\sin^2(q_x)+\sin^2(q_y)) \nonumber\\
 &&- L_{{\mathbf q}} (2(\cos 2q_x+\cos 2q_y)-2)\Bigr) \nonumber\\
&&+J_2 \Bigl(K_{{\mathbf q}}(\mp 4\cos(q_x)\cos (q_y)\nonumber\\&&
\pm 2+2\cos (q_x+q_y))\nonumber\\
&&-L_{{\mathbf q}} (2\cos(q_x-q_y)\mp 2)\Bigr),\nonumber\\
 B^{(4)}_\mathbf{Q_{1(2)}}&=& \frac{1}{4N}\sum_\mathbf{q} J_3
\Bigl( K_{\mathbf q}(
\cos 2q_x+\cos 2q_y-4)\nonumber\\
&&+2L_{\mathbf q} (\cos 2q_x+\cos2q_y)\Bigr )\nonumber\\
&&+J_2 \Bigl( K_{{\mathbf q}} (\cos(q_x-q_y)\mp 2)\nonumber\\
&&+L_{\mathbf q}(\pm 4\cos (q_x)\cos (q_y)\nonumber\\&&
\mp 1 -\cos (q_x+q_y))\Bigr)~,\nonumber
\end{eqnarray}
where the upper  and the lower signs correspond to $\mathbf{Q_{1}}$ and $\mathbf{Q_{2}}$ vectors, respectively.
 One can easily make sure that $A^{(4)}_\mathbf{Q_{1}}=B^{(4)}_\mathbf{Q_{1}}$, but $A^{(4)}_\mathbf{Q_{2}} \neq B^{(4)}_\mathbf{Q_{2}}$.  Thus, when 1/S corrections are included, the spectrum at $\mathbf{Q_{1}}$ remains gapless, as it should be,
   but at $\mathbf{Q_{2}}$  the gap opens up, i.e., quantum fluctuations lift the accidental degeneracy at the "wrong $Q$".   We computed the renormalized spectrum for all ${\mathbf k}$ and show the results in  Fig.3. The renormalized spectrum, shown in the bottom panel, clearly has a gap at $\pm (\pi/2,\pi/2)$, where the spectrum of non-interacting bosons (top panel) has zero modes.
   We verified that the same effect (lifting of accidental nodes) holds if we use biquadratic spin interaction instead of quantum fluctuations.

\title{{\it Supplemental Material for }  Magnetism in
 parent Fe-chalcogenides: quantum fluctuations select a plaquette order}
\author{Samuel Ducatman, Natalia  B. Perkins, Andrey Chubukov}
\affiliation{Department of Physics, University of Wisconsin-Madison, Madison, WI 53706, USA}

\maketitle

 In this Supplementary Material we provide details  of  our  analysis of how quantum fluctuations lift  accidental nodes
  in the spin-wave spectra
 of the sub-set of states selected by quantum fluctuations
   in the  $J_1-J_2-J_3$ model at $J_3 > J_2/2$ and $J_1=0$ (i.e., in the $J_2-J_3$ model with $J_3 > J_2/2$).
    The Hamiltonian for the model is given by Eq.~(1) in the main text.  At $J_1 =0$ spins on
   even and  odd sites do not interact with each other. It is therefore sufficient to consider
   only the sub-set made out of spins on, say, even sites.

 Classically degenerate ground states are shown in Fig.~2 of the main text.  For spins on even sites, any two-sublattice
 configuration with arbitrary angle $\gamma$ between spins in $a$ and $b$
  sublattices is the ground state. Like we said in the text, quantum fluctuations break this degeneracy and select the states at $\gamma =0$ or $\gamma = \pi$ as the only two ground states.  The state with $\gamma =0$ can be viewed as a
   stripe
   state with $Q_1 = (\pi/2, -\pi/2)$, and the $\gamma =\pi$ state corresponds to a
    stripe
     with $Q_2 = (\pi/2, \pi/2)$.
   Because only even cites are involved, configurations with $Q_i$ and $-Q_i$ are identical.

  The classical spectrum of either $Q_1$ or $Q_2$ state contains zero modes at corresponding $Q$, which must be there by Goldstone theorem, but also contains zero modes at the "wrong Q"  (at $Q_2$ for the $Q_1$
    stripe
     and vise versa). These last modes are not associated with symmetry breaking (i.e., are "accidental") and must be lifted by quantum fluctuations.  We argued in the
      main text that quantum fluctuations as well as biquadratic exchange coupling  gap accidental zero modes,  and studied the consequences.  Here we show how this actually happens in the quantum case. For definiteness, we consider $Q_1$ configuration ($\gamma =0$).

\begin{figure}
\includegraphics[width=0.85\columnwidth]{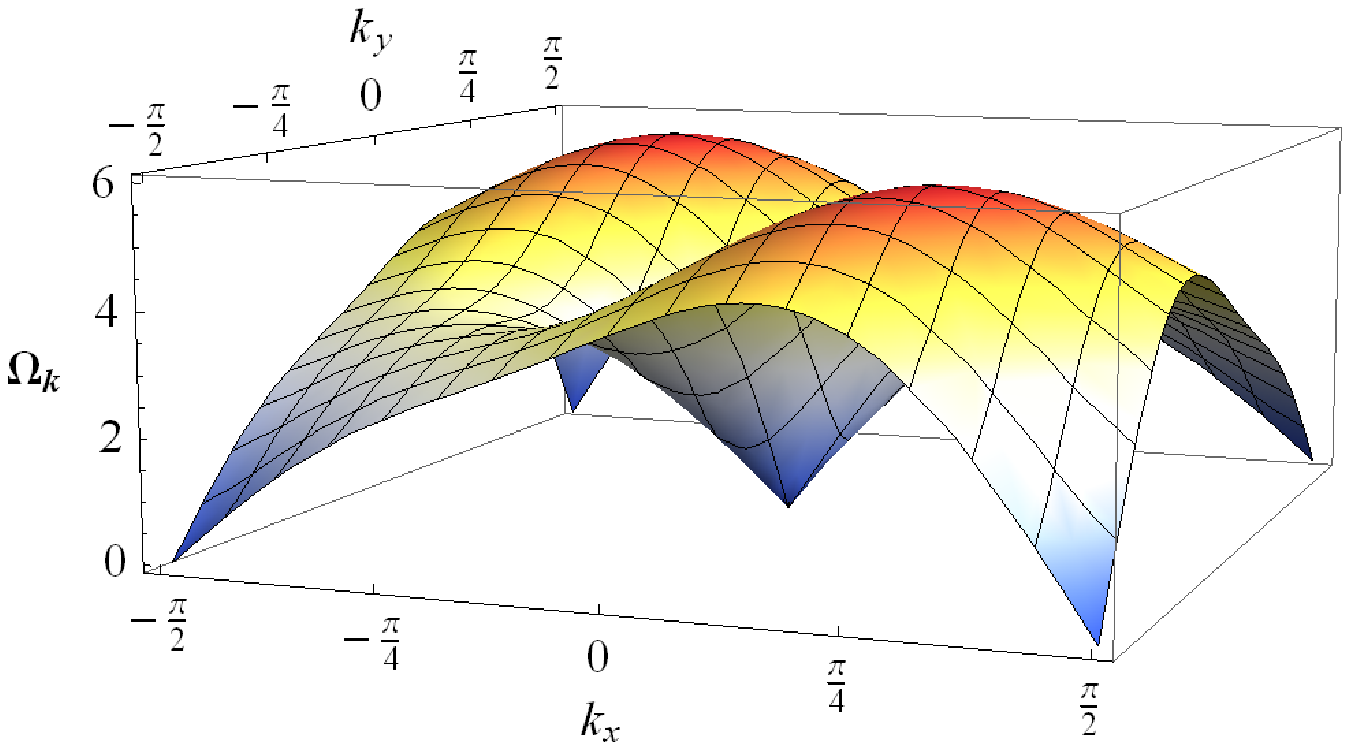}
\includegraphics[width=0.99\columnwidth]{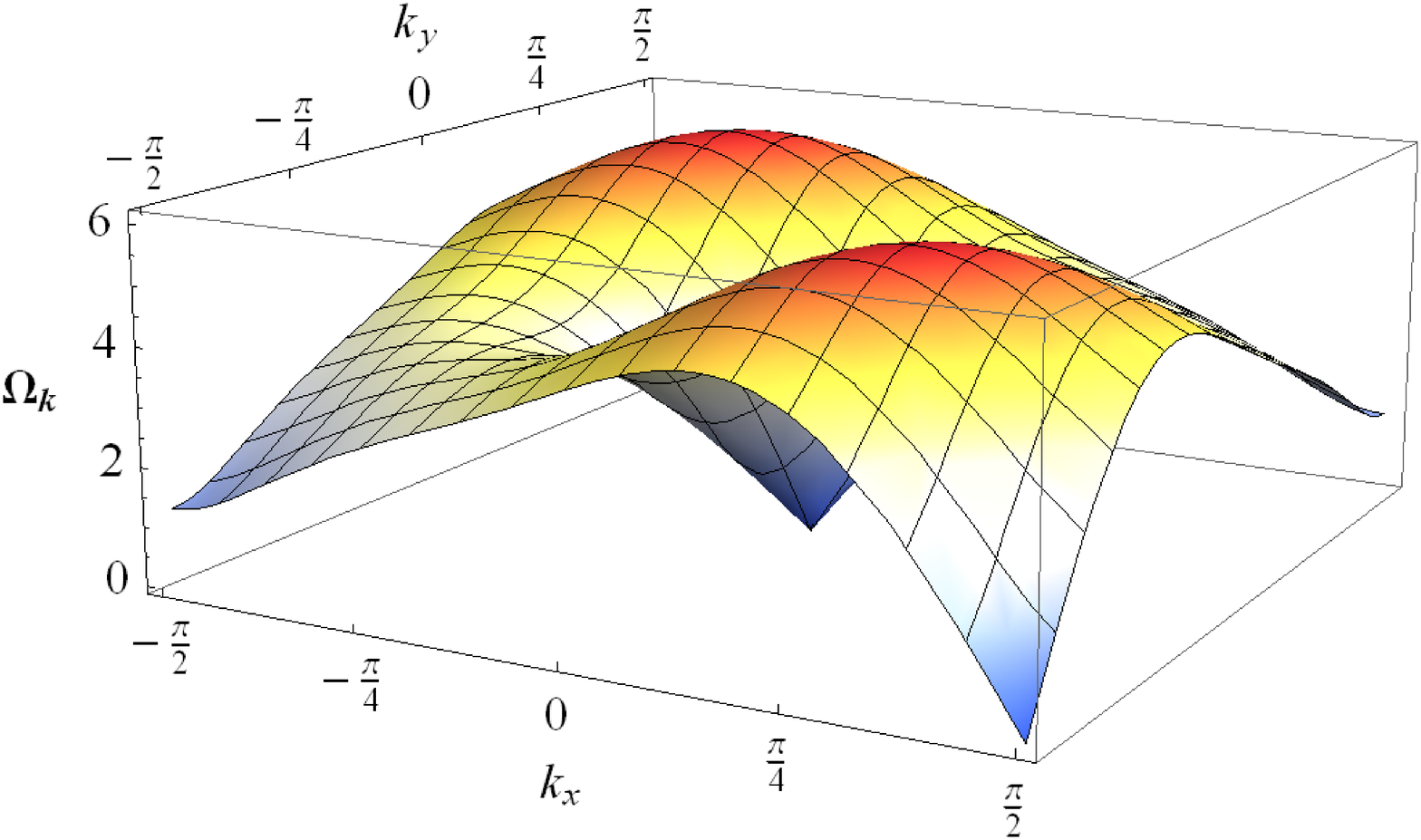}
\caption{Excitation spectrum for $Q_1$
 stripe.
 Top panel: The classical spectrum of $J_2-J_3$ model  computed for $J_2=J_3=1$.
 This spectrum contains zero modes at $\pm (\pi/2, \pi/2)$ and $\pm (\pi/2, -\pi/2)$.
  Bottom panel: the  spectrum for $J_2-J_3$ model in the presence of biquadratic exchange coupling $K=0.05$.
 The true Goldstone modes at
$\pm\mathbf{Q_{1}}=\pm(\frac{\pi}{2},-\frac{\pi}{2})$ remain, but  the  accidental
zeroes at $\pm \mathbf{Q}_2=\pm(\frac{\pi}{2},\frac{\pi}{2})$   are
removed.}\label{fig:spectrum}
\end{figure}

The lifting of accidental zero modes can be studied within the original two-sublattice picture, with two different bosons describing fluctuations of spins in $a$ and $b$ sublattices. It is more convenient, however,
 to  perform a uniform rotation
  to a local reference frame in which the magnetic order is ferromagnetic, and describe the excitations  using  one
  Holstein-Primakoff  bose operator $a_i$:
 \begin{eqnarray}
    \label{eq:HP}\nonumber
    \begin{array}{l}
    S^z_i =S - a^{\dagger}_i a_i \\
    S^+_i =(2S - a^{\dagger}_i a_i)^{\frac{1}{2}}a_i~~~~~~~~~~(S.1) \\
    S^-_i = a^{\dagger}_i (2S -a^{\dagger}_ia_i)^{\frac{1}{2}}~,
    \end{array}
\end{eqnarray}

As it is customary to spin-wave analysis, we assume that $S$ is large.
Our goal is to obtain the excitation spectrum for $Q_1$ configuration including terms
 of order $1/S$, which describe quantum fluctuations.
To get this spectrum,  we  substitute Eq.~(S.1) into
Hamiltonian (Eq.~(1)) and expand  it to the quartic order in $a$ field.
 We obtain
 \begin{eqnarray}
    \label{eq:Hb}\nonumber
    H = E_0+S(H_2+H_4)~,~~~~~~~(S.2)
\end{eqnarray}
where $H_2$ and $H_4$ are the quadratic and the quartic terms, respectively.
In momentum space, we have
 \begin{eqnarray}
    \label{eq:H2}
\nonumber
H_2 =& \sum_\mathbf{k} \Bigl( A_\mathbf{k}a^{\dagger}_\mathbf{k}a_\mathbf{k}
 +\frac{B_\mathbf{k}}{2}(a_\mathbf{k}a_\mathbf{-k}+h.c.)
\Bigr ),~~(S.3)
\end{eqnarray}
where
\begin{eqnarray}\nonumber
A_\mathbf{k}&=& 4J_3 +2J_2\cos(k_x+k_y) ~~~~~~~~~~~~~~~(S.4)  \\\nonumber
B_\mathbf{k}&=& -2J_3(\cos(2k_x)+\cos(2k_y))-2J_2 \cos(k_x-k_y)
\end{eqnarray}
and ${\bf k}$ is defined in the first magnetic BZ.

The quadratic Hamiltonian $H_2$ is diagonalized
 by the Bogoliubov
transformation, $\alpha_\mathbf{k} = u_\mathbf{k}a_\mathbf{k}-v_\mathbf{k}a^{\dagger}_\mathbf{-k}$,
where  the  factors $u_\mathbf{k}$ and $v_\mathbf{k}$ are determined by
 \begin{eqnarray}
\nonumber
u_\mathbf{k} &=& \sqrt{\frac{A_\mathbf{k}+\Omega_{\mathbf{k}}}{2\Omega_{\mathbf{k}}}}~, \\
v_\mathbf{k}&=& -\text{sign}B_\mathbf{k}\sqrt{\frac{A_\mathbf{k}-\Omega_{\mathbf{k}}}
{2\Omega_{\mathbf{k}}}}\nonumber
\end{eqnarray}
 and
\begin{eqnarray}
\Omega_{\mathbf{k}}= S\sqrt{A_\mathbf{k}^2 - B_\mathbf{k}^2}~.
\label{y_1}
\end{eqnarray}
The diagonalized Hamiltonian $H_2$ is  given by
\begin{eqnarray}\nonumber
H_2=E_{2} + \sum_\mathbf{k} \Omega_{\mathbf{k}}\alpha^\dagger_\mathbf{k} \alpha_\mathbf{k}~,~~~~~~~~(S.5)
\end{eqnarray}
where
\begin{eqnarray}
    \label{e2}
E_{2}=\frac{S}{2}\sum_{\mathbf k} (-A_\mathbf{k} + \Omega_{\mathbf{k}})~~~~~~~~~~~~(S.6)\nonumber
\end{eqnarray}
 is the  contribution to the ground state energy from non-interacting bosons

 Eq. (\ref{y_1}) gives spin-wave dispersion without $1/S$ corrections.
  At  $\mathbf{k} = \pm Q_1$ and $\mathbf{k} = \pm Q_2$, $A_{Q_{1(2)}}=B_{Q_{1(2)}}=4J_3 \pm 2J_2$, where  + and - signs are  for  $Q_1$ and $Q_2$, respectively.  As a result, $\Omega_{Q_1}=\Omega_{Q_2}=0$.  However, only the zero mode at $\mathbf{k} = \pm Q_1$ is the
   true Goldstone mode,  the other one is accidental and must be lifted by quantum fluctuations.
   To show this, we
  compute 1/S corrections  to the spectrum at these points.

The quartic  term in the Hamiltonian is  given by
\begin{eqnarray}\label{eq:H4}
&&H_4 =\frac{1}{2S}\sum_{ij} J_{ij}\Bigl(\cos\phi_{ij}a^{\dagger}_ia_ia^{\dagger}_ja_j+~~~~~~(S.7)\nonumber\\
&&\frac{1-\cos\phi_{ij}}{8}\bigl(a_ia^{\dagger}_ja_ja_j+
a^{\dagger}_ia_ia_ia_j+a^{\dagger}_ia^{\dagger}_ja^{\dagger}_ja_j+a^{\dagger}_ia^{\dagger}_ia_ia^{\dagger}_j\bigr)-
\nonumber\\\nonumber
&&\frac{1+\cos\phi_{ij}}{8}
\bigl(a_ia^{\dagger}_ja^{\dagger}_ja_j+a^{\dagger}_ia_ia_ia^{\dagger}_j+a^{\dagger}_ia^{\dagger}_ja_ja_j+
a^{\dagger}_ia^{\dagger}_ia_ia_j\bigr)\Bigr)
\end{eqnarray}
 The angle $\phi_{ij}=0$ for second neighbors along one diagonal and $\pi$ along the other. For  third neighbors, $\phi_{ij}=\pi$ for all neighbors.
  Performing Fourier transformation to the momentum space, we obtain
 \begin{eqnarray}\label{eq:H41}
&&H_4 = \frac{J_2}{NS}
\sum_{\{\mathbf k_i\}}\delta\Bigl(\sum_i\mathbf k_i \Bigr)~~~~~~~~~(S.8)\nonumber\\\nonumber
&&\Bigl[
a^{\dagger}_{{\mathbf k}_1}a^{\dagger}_{{\mathbf k}_2}a_{{\mathbf k}_3}a_{{\mathbf k}_4}
(\cos(k_{2x}-k_{4x}+k_{2y}-k_{4y})-\\\nonumber
&&\cos(k_{2x}-k_{4x}-k_{2y}+k_{4y})-
\frac{1}{4}(\cos(k_{1x}+k_{1y})+\\\nonumber
&&\cos(k_{2x}+k_{2y})+
\cos(k_{3x}+k_{3y})+\cos(k_{4x}+k_{4y})))+\\\nonumber
&&
\frac{1}{4}(
a^{\dagger}_{{\mathbf k}_1}a_{{\mathbf k}_2}a_{{\mathbf k}_3}a_{{\mathbf k}_4}(\cos(k_{2x}-k_{2y})+\cos(k_{4x}-k_{4y}))+\\\nonumber
&&a^{\dagger}_{{\mathbf k}_1}a^{\dagger}_{{\mathbf k}_2}a^{\dagger}_{{\mathbf k}_3}a_{{\mathbf k}_4}(\cos(k_{1x}-k_{1y})+\cos(k_{3x}-k_{3y})))
\Bigr]
\end{eqnarray}

The $1/S$ contribution to $\Omega_{\mathbf{k}}$ is obtained by evaluating first-order correction to spin-wave dispersion from $H_4$.
  For this, we decouple each term in $H_4$ into the product of two pairs of bosons and replace one pair by its average value for non-interacting bosons.
   The average values are
\begin{eqnarray}
K_{{\mathbf q}} &=& \langle a_{\mathbf q}^{\dagger} a_{\mathbf q} \rangle  = v_{\mathbf q}^2 ~,\nonumber\\
L_{{\mathbf q}} &=& \langle a_{\mathbf q} a_{-{\mathbf q}} \rangle = \langle a_{-\mathbf q}^{\dagger} a_{{\mathbf q}}^{\dagger}  \rangle = u_{\mathbf q}v_{\mathbf q} ~.\nonumber
\end{eqnarray}
The decoupled $H_4$ is then
\begin{eqnarray}
{\bar H}_4 &=&\frac{1}{NS} \sum_{\mathbf k}
\Bigl(a^{\dagger}_{\mathbf k} a_{\mathbf k} A^{(4)}_\mathbf{k}+\nonumber\\&&
\frac{1}{2}(a_{\mathbf k} a_{-{\mathbf k}} +a_{-\mathbf k}^{\dagger} a_{{\mathbf k}}^{\dagger}  )
B^{(4)}_\mathbf{k}\Bigr),~(S.9)\nonumber
\end{eqnarray}
where
\begin{eqnarray}
&&A^{(4)}_\mathbf{k} = J_2\sum_{\mathbf q}\Bigl(-2K_{{\mathbf q}}(\cos((k_x-k_y)-(q_x-q_y))-\nonumber\\\nonumber
&& \cos((k_x+k_y)-(q_x+q_y))+\cos(k_x+k_y)+\cos(q_x+q_y))\\\nonumber
&& +L_{{\mathbf q}}(2\cos(q_x-q_y)+\cos(k_x-k_y))\Bigr)+\\\nonumber
&& J_3\sum_{\mathbf q}\Bigl(-4K_{{\mathbf q}}-2K_{{\mathbf q}}(\cos2(k_x-q_x)+\cos2(k_y-q_y))+\\\nonumber
&&L_{{\mathbf q}}(2\cos2q_x+2\cos2q_y+\cos2k_x+\cos2k_y)\Bigr)
\end{eqnarray}
and
\begin{eqnarray}
&&B^{(4)}_\mathbf{k} = J_2\sum_{\mathbf q}\Bigl(
-2L_{{\mathbf q}}\cos((k_x-k_y)-(q_x-q_y))+\nonumber\\\nonumber
&& 2L_{{\mathbf q}}\cos((k_x+k_y)-(q_x+q_y))+ \\\nonumber&& K_{{\mathbf q}}(2\cos(k_x-k_y)+\cos(q_x-q_y))-\\\nonumber
&&L_{{\mathbf q}}(2\cos(k_x+k_y)+\cos(q_x+q_y))\Bigr)\\\nonumber
&& +J_3\sum_{\mathbf q}\Bigl(-2L_{{\mathbf q}}(\cos2(k_x-q_x)+\cos2(k_y-q_y))
+\\\nonumber&&
K_{{\mathbf q}}(2\cos 2k_x+2\cos 2 k_y+\cos2q_x+\cos2q_y)\Bigr)
\end{eqnarray}

Combining the  decoupled $H_4$ with $H_2$, expressed in terms of original $a$-operators, we obtain
\begin{eqnarray}\label{h4deca}
    &&H_2 + {\bar H}_4
    =   \sum_{\mathbf k} \Bigl[ (A_\mathbf{k}+\frac{A^{(4)}_\mathbf{k}}{S})a^{\dagger}_\mathbf{k} a_\mathbf{k}
     + \nonumber\\&&
  \frac{1}{2}(B_\mathbf{k}+\frac{B^{(4)}_\mathbf{k}}{S})(a_\mathbf{k}a_\mathbf{-k}+h.c.)\Bigr],~~~~(S.10)\nonumber
\end{eqnarray}

At $\mathbf{k} = Q_1$
\begin{eqnarray}
&&A^{(4)}_\mathbf{Q_{1}}=
J_2\sum_{\mathbf q}\Bigl(-2K_{{\mathbf q}}-L_{{\mathbf q}}+\nonumber\\\nonumber
&&2(K_{{\mathbf q}}+L_{{\mathbf q}})\cos (q_x-q_y)\Bigr)+
J_3\sum_{\mathbf q}\Bigl(-4K_{{\mathbf q}}-\\\nonumber
&&2L_{{\mathbf q}}+2(K_{{\mathbf q}}+L_{{\mathbf q}})(\cos2q_x+\cos2q_y)\Bigr)
\end{eqnarray}
\begin{eqnarray}
&&B^{(4)}_\mathbf{Q_{1}}=
J_2\sum_{\mathbf q}\Bigl(-2K_{{\mathbf q}}-2L_{{\mathbf q}}+
(K_{{\mathbf q}}+2L_{{\mathbf q}})\cos (q_x-q_y)\Bigr)\nonumber\\\nonumber&&
+J_3\sum_{\mathbf q}\Bigl(-4K_{{\mathbf q}}+(K_{{\mathbf q}}+2L_{{\mathbf q}})(\cos2q_x+\cos2q_y)\Bigr),
\end{eqnarray}
and at $\mathbf{k} = Q_2$
\begin{eqnarray}
&&A^{(4)}_\mathbf{Q_{2}}=J_2\sum_{\mathbf q}\Bigl(2K_{{\mathbf q}}+
L_{{\mathbf q}}-2(K_{{\mathbf q}}-L_{{\mathbf q}})\cos (q_x-q_y)\nonumber\\\nonumber&&
-4K_{{\mathbf q}}\cos (q_x+q_y)\Bigr)+
J_3\sum_{\mathbf q}\Bigl(-4K_{{\mathbf q}}-\\\nonumber
&&2L_{{\mathbf q}}+2(K_{{\mathbf q}}+L_{{\mathbf q}})(\cos2q_x+\cos2q_y)\Bigr)
\end{eqnarray}
\begin{eqnarray}
&&B^{(4)}_\mathbf{Q_{2}}=
J_2\sum_{\mathbf q}\Bigl(2K_{{\mathbf q}}+2L_{{\mathbf q}}+(K_{{\mathbf q}}-2L_{{\mathbf q}})\cos (q_x-q_y)\nonumber\\\nonumber
&&-2L_{{\mathbf q}}\cos (q_x+q_y)\Bigr)\\\nonumber&&
+J_3\sum_{\mathbf q}\Bigl(-4K_{{\mathbf q}}+(K_{{\mathbf q}}+2L_{{\mathbf q}})(\cos2q_x+\cos2q_y)\Bigr).
\end{eqnarray}
 One can easily make sure that
 $A^{(4)}_\mathbf{Q_{1}}=B^{(4)}_\mathbf{Q_{1}}$, hence the spectrum at $\mathbf{Q_{1}}$ remains gapless even with $1/S$ corrections, as it should be because
  $Q_1$ is a true Goldstone point. At the same time, at $\mathbf{Q_{2}}$,  $A^{(4)}_\mathbf{Q_{2}} \neq B^{(4)}_\mathbf{Q_{2}}$.   As a result,
   $1/S$ corrections open up the gap in the spin-wave spectrum at $\mathbf{Q_{2}}$.  We computed this gap numerically and used the result in the main text.

   The same result is obtained if we add biquadratic coupling $K \left({\bf S}_i {\bf S}_j\right)^2$ between second neighbors. We do not present the details of the calculations (they are straightforward) but show the result in Fig.1.

\end{document}